\documentclass[aip,cha,numerical,preprint]{revtex4-1}

\usepackage[dvipdfmx]{graphicx}%
\usepackage{dcolumn}%
\usepackage{bm}%
\draft 

\begin{document}


\title{Dynamics of impurity attraction and repulsion of an intrinsic localized mode in a driven 1-D cantilever array} 



\author{M. Sato}
\email[]{msato153@staff.kanazawa-u.ac.jp}
\author{Y. Sada}
\author{W. Shi}
\author{S. Shige}
\author{T. Ishikawa}
\author{Y. Soga}
\affiliation{Graduate School of Natural Science and Technology, Kanazawa University\\Kanazawa, Ishikawa 920-1192, Japan}
\author{B. E. Hubbard}
\affiliation{Laboratory of Atomic and Solid State Physics, Cornell University\\
Ithaca, NY 14853-2501, USA}
\author{B. Ilic}
\affiliation{Cornell Nanoscale Science and Technology Facility, Cornell University\\Ithaca, NY 14853-2501, USA}
\author{A. J. Sievers}
\affiliation{Laboratory of Atomic and Solid State Physics, Cornell University\\
Ithaca, NY 14853-2501, USA}

\date{\today}

\begin{abstract}
Both low frequency and high frequency impurity modes have been produced in a SiN micromechanical cantilever array by illumination with either an infrared or visible laser. When such laser-induced impurities are placed near a driven intrinsic localized mode (ILM) it is either repelled or attracted. By measuring the linear response spectrum for these two cases it was found that vibrational hopping of the ILM takes place when the natural frequency of the ILM and an even symmetry linear local mode are symmetrically located about the driven ILM frequency so that parametric excitation of these two linear modes is enhanced, amplifying the lateral motion of the ILM. Numerical simulations are consistent with these signature findings. It is also demonstrated that the correct sign of the observed interaction can be found with a harmonic lattice-impurity model but the magnitude of the effect is enhanced in a nonlinear lattice.
\end{abstract}


\maketitle 

\begin{quotation}
An intrinsic localized mode (ILM) represents a localized vibrational excitation in a nonlinear lattice. Such a mode will stay in resonance as the driver frequency is changed until the ILM switches and disappears. The dynamical properties of this robust excitation can be examined by tickling it in various ways to see how it responds. Such linear response spectra have been measured for a driven ILM in a micromechanical array containing an impurity. It has been shown that a visible laser-induced impurity will attract the ILM while an IR laser-induced impurity will repel it. The reason for these experiments is to explore the dynamics associated with the transition from stationary to moving ILM. We find in both cases that the ILM transition occurs when the frequencies of two different linear modes, produced by the ILM, are located symmetrically about its driven frequency. Since this arrangement ensures the maximum nonlinear transfer of energy between the ILM and its translational component these findings are expected to be quite general for a driven 1-D nonlinear lattice.
\end{quotation}

\section{Introduction}
Because of the many applications of laser beam deflection measurement of a microcantilever a number of studies have explored how the laser spot may change the mechanical properties of the cantilever\cite{1,2,3,4} and affect the accuracy of its vibrational frequency measurement.\cite{5,6} It is known that Joule heating, via both a temperature dependent Young's modulus and thermal expansion, shifts the vibrational frequency to the red for silicon and silicon nitride.\cite{7,8} At low laser power the expected blue shift from radiation-induced pressure change in the vibrational frequency is usually too small to be important.\cite{5} Recently we have observed a visible laser-induced blue frequency shift for the vibrational frequency of SiN cantilevers, which may be the result of a surface stress generated by photo-excited carriers. This last experimental observation has opened up the opportunity to quantify a nonlinear dynamical effect that was reported some years back.\cite{9,10} 

The earlier discovery that localized dynamical energy can be stabilized in a perfect nonlinear lattice by its discreteness had demonstrated that both extended and localized waves were necessary features of such systems. \cite{Dolgov,ST,FG}  One end result was that dynamically driven intrinsic localized modes (ILMs) provided a new window into the nonlinear dynamics structure of discrete systems. \cite{12} Our qualitative observation for a micromechanical array was that when heat from a focused IR laser was applied to a cantilever near a driven ILM on a 1-D lattice the ILM would move to the next lattice site, away from the laser spot, if the lattice had hard nonlinearity (with increasing amplitude the frequency increases); while, if the lattice had soft nonlinearity (the converse) the ILM would hop toward the IR laser spot. \cite{9,10,EPL7} With sufficient laser power the ILM would track the moving laser spot. \cite{EPL7} The recent discovery of the visible laser-induced blue shift of the cantilever resonance in SiN now makes possible a detailed investigation of the dynamics behind the IR laser-induced pushing or visible laser pulling of the ILM in the same lattice.  

Experimentally, a stationary ILM can be maintained in steady state using a driver to compensate for damping.\cite{9,11,12,13}  In this case the ILM is frequency locked, and an auto-resonance (AR) state \cite{14,15,16,17,18,19} is achieved where the driver frequency controls the ILM amplitude. Previously, it has been demonstrated that a linear response measurement in the driven 1-D array is a powerful tool for studying bifurcation dynamics of an ILM.\cite{20,21,22} 

Here this same technique is used to follow both the release and capture of an ILM by laser generated impurities. With application of the IR laser the soft impurity produces a nonlinear coupling between the natural frequency (NF) of the ILM and a nearby even linear local mode (LLM)~\cite{23} associated with the ILM, giving rise to parametric amplification of both of these modes. Sufficient amplitude can be generated to induce translational ILM motion. When the IR laser spot is replaced by a visible one so that a hard impurity is generated the same mixing process again provides the necessary translational amplitude. The sign of the translation is controlled by the frequency of the impurity mode relative to the plane wave band. Our numerical simulations indicate that when it is a soft impurity mode the center of gravity of the ILM moves away from the impurity while the opposite occurs for a hard impurity mode.

The next section illustrates the experimental setup and describes the measurement procedure for the micromechanical array. Section~\ref{sec:experiment} first presents the experimental results for the visible laser hardening of a single vibrating cantilever and then shows the linear response measurements for the driven nonlinear di-element lattice using a pump-probe method while either an infrared or visible laser generates a soft or hard impurity, respectively. In section~\ref{sec:simulation} simulation results are compared with experiments. It is verified that the transition from stationary to moving ILM occurs when the natural frequency (NF) and the even-LLM frequencies are symmetrically located about the driven ILM frequency. It is also demonstrated with a harmonic impurity-impurity model that the sign of the effect is consistent with the observed impurity-induced repulsion and attraction. The mechanism of cantilever hardening is discussed and the conclusions follow.

\section{Experimental method and procedure}
\subsection{Experimental arrangement}
\begin{figure}
\includegraphics{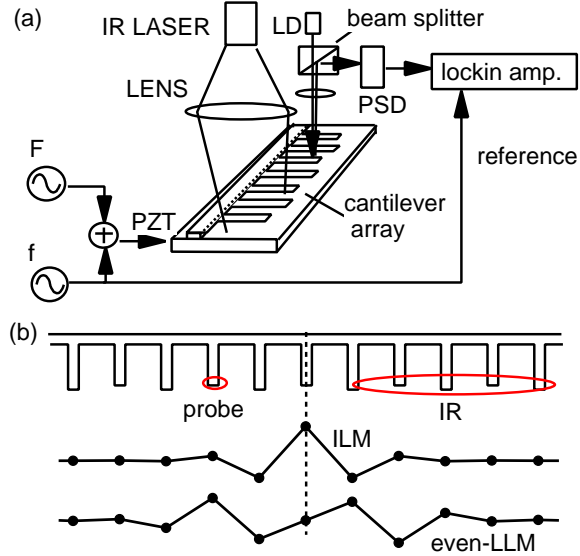}%
\caption{\label{fig:1}(a) Experimental set up for the linear response measurement of the autoresonant ILM and a laser induced impurity mode. Array is composed of SiN cantilevers of two lengths,  50 and 55 $\mu$m aligned alternatively. Substrate is a silicon wafer. The pump driver at frequency $F$ and probe signal at frequency $f$ are added and used to excite the array uniformly via the thin PZT. A low power visible laser diode (LD) illuminates a cantilever nearby the ILM and the reflected beam is detected by a position sensitive detector (PSD). For ILM repulsion the IR laser beam is soft focused at a side of the ILM. For ILM attraction the visible laser beam with increased power is focused on a single cantilever next to the ILM. The displacement signal is analyzed by a lock-in amplifier. Typical pump amplitude is 14V, while the probe amplitude is 12mV. (b) Spatial pattern of the ILM and associated even mode LLM. The two ellipses indicate sizes of laser beams for probe detection and local heating for the repulsion case. For attraction case the IR laser is removed and the probe laser power increased.}%
\end{figure}

Figure~\ref{fig:1}(a) shows the experimental setup. The driven micromechanical array contains 152 di-element SiN cantilevers coupled together by the common overhang on a silicon wafer. Because of positive nonlinearity of the cantilever array, the ILM is generated above the top of the linear dispersion curve. A di-element array was used so that the highest frequency normal mode could be excited by the application of a uniform driver, which is a piezoelectric transducer (PZT). Because of the mass difference between short and long cantilevers different forces are felt from the uniform acceleration. A cw pump oscillator feeds energy to the array maintaining the ILM in the large amplitude AR state. Generation and movements of ILMs were monitored by  a camera [not shown in Fig. 1(a)]. By increasing the driver frequency above the top of the dispersion curve, ILMs can be generated. For the IR laser repulsion case, the IR laser beam is placed to one side of the ILM and then the IR laser power is increased until the ILM moves away from the laser spot. A visible laser at a low power level (0.06 mW) is used to monitor the motion of one cantilever at the other side of the IR laser spot as shown in Fig. 1(b). The driver frequency is fixed during this transition measurement process. For the visible laser attraction case, a higher power (1 mW) visible laser is used. Then, the laser beam is placed to one side of the ILM and the driver frequency decreased until the ILM is attracted to the laser beam. The reflection from the visible laser beam is used to monitor the motion of the cantilever.

For linear response measurements an additional weak probe driver of variable frequency is combined with the strong pump so the perturbation is applied uniformly across the lattice. With an ILM present the motion of a nearby single cantilever is monitored using the visible laser and a position sensitive detector (PSD, S3932 Hamamatsu Photonics). A lock-in amplifier selectively detects the vibration that is caused by the probe oscillating at a given frequency. A response spectrum is measured by scanning the probe frequency, while other conditions, such as the pump frequency and laser power are held fixed. Vibrational modes are recorded as either positive or negative peaks in the imaginary part of the response spectra.

\subsection{Localized vibrational modes in the linear experiments}
The spectrum of linear local modes associated with a large amplitude ILM is now fairly well understood.\cite{21,23} One mode is the natural frequency (NF) of the ILM. This NF is for an ILM with the same vibrational amplitude pattern but without driver or damping. The driven ILM also gives rise to nearby LLMs.

With a fixed boundary condition, or by symmetry breaking by the AR state, all linear modes can be classified into odd and even symmetries in terms of the vibrating spatial pattern and appear alternatively when ordered by their mode frequency. For example, the ILM shown in Fig.~\ref{fig:1}(b) is odd and the LLM is even. Usually our driving method couples to the odd mode vibrational pattern (... -0.4, 1, -0.4, ...) that have a finite coupling to the uniform acceleration of the PZT with the di-element mass pattern (... $m_a$, $m_b$, $m_a$, ...). Normally even modes are not recorded in these spectra but if an impurity exists at one side of the ILM, the symmetry is broken and the even mode becomes observable in our experiment.

\section{\label{sec:experiment}Experimental results}
\subsection{Single cantilever-visible laser results}
\begin{figure}
\includegraphics{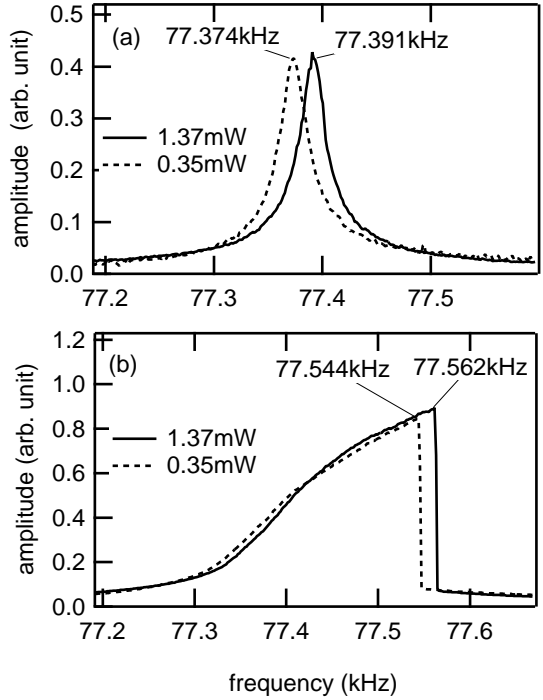}%
\caption{\label{fig:2}Effect of the visible laser on a single cantilever at two different powers, 0.35 and 1.37 mW.  Cantilever size: 300nm thick, 15$\mu $m width and 60$\mu $m long. (a) Linear resonance peak is shifted from 77.374kHz to 77.391 kHz with increased power. (b) Nonlinear response curve with a larger pump amplitude is also changed slightly, at the upper bifurcation frequency of the Duffing response from 77.544kHz to 77.562 kHz.}%
\end{figure}

Since laser radiation is used to produce the impurity we have checked the influence of the radiation on the lowest frequency mode response curve of a single cantilever  made from the same SiN material as the array as a function of different laser powers. For the IR laser the effect is simply to soften the linear spring constant so that the cantilever frequency decreases. The visible laser result for different laser powers is more interesting. Figure~\ref{fig:2} shows both the linear and nonlinear response curves versus visible laser power. In both cases, the response curves are shifted upwards in frequency as the laser power increases. Since both curves shift the same way, we can conclude that the application of visible laser radiation increases the linear resonance frequency. This shift can be converted into an increase of a harmonic spring constant (assuming a ball and spring model.) From the equation $\Delta k/k=2\Delta f/f$, the increment is estimated as 0.04\%.

\subsection{IR laser-induced ILM repulsion experiment for the array}
\begin{figure}
\includegraphics{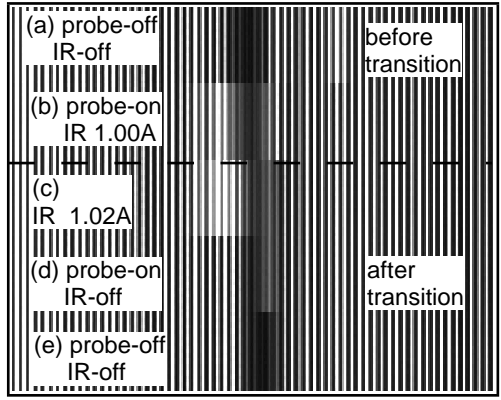}%
\caption{\label{fig:3}CCD camera images for the IR laser ILM-impurity mode repulsive transition. Vertical white lines are cantilevers. The five images taken at different time are aligned vertically. Fixed driver frequency.  (a) ILM without probe and IR lasers. The ILM is identified by dark cantilevers because of their large vibration amplitudes. Size of the ILM is 5 cantilevers, which is about 3 unit cells. (b) The probe laser beam is located at the right side of the ILM.  The IR laser beam is set at the left side with a larger spot size (white region). (c) Repulsive hopping transition takes place when the IR laser current is increased to 1.02A. (d) The ILM remains at the new position after turning off the IR laser. (e) Without the probe beam the ILM pattern is the same as case (a) but displaced.}%
\end{figure}

The display in Fig.~\ref{fig:3}(a) shows the initial condition for the repulsion measurement. First, the soft-focused IR laser beam at very low power level is placed to one side of the ILM. The probe laser is placed on the opposite side of the ILM with respect to the IR laser, as shown in Fig.~\ref{fig:3}(b). By changing the IR laser current in a stepwise fashion, the linear mode properties can be monitored as the hopping transition is approached.  Measurement of the spectrum and step increment of the IR laser current are repeated alternatively till the ILM is repelled by the IR laser as shown in Fig.~\ref{fig:3}(c).

\begin{figure}
\includegraphics{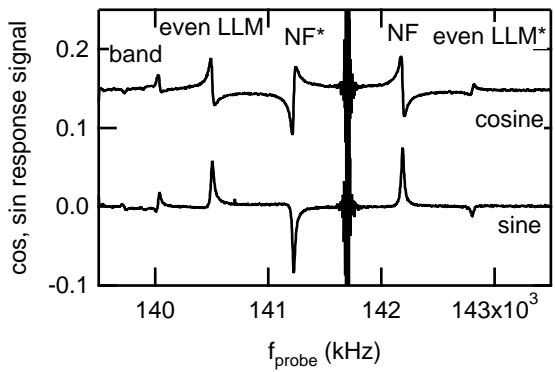}%
\caption{\label{fig:4}Linear response curves for the soft impurity ILM interaction. Cosine (upper) and sine (lower) outputs of the lock-in amplifier recorded as a function of the probe frequency. The pump driver frequency $F$ (ILM frequency) was fixed at 141.7 kHz, which is slightly larger than the band top frequency of the array (140 kHz).  The bandwidth of the upper branch ($f_{BW}$) is 3 kHz. Two sets of positive and negative peaks in the sine response are observed as sidebands. NF and even-LLM are identified as positive peaks on the high frequency side and the low frequency side, respectively. Negative peaks are due to four-wave mixing of corresponding mode and the ILM. These partner modes are identified with "*". The even LLM is activated by the IR laser beam placed a side of the ILM. IR laser current is 0.98A.}%
\end{figure}

Figure~\ref{fig:4} shows the typical cosine and sine linear response spectra versus probe frequency. The central peak is the large amplitude ILM vibration. The upper positive peak is the NF of the ILM. The NF is accompanied by its four-wave mixing partner NF$^*$, showing a negative peak in the sine spectrum on the low frequency side, at the same distance from the central peak. In addition to the NF and band modes, the even LLM is identified. The soft impurity is to one side of the ILM where the even LLM signal is larger.

\begin{figure}
\includegraphics{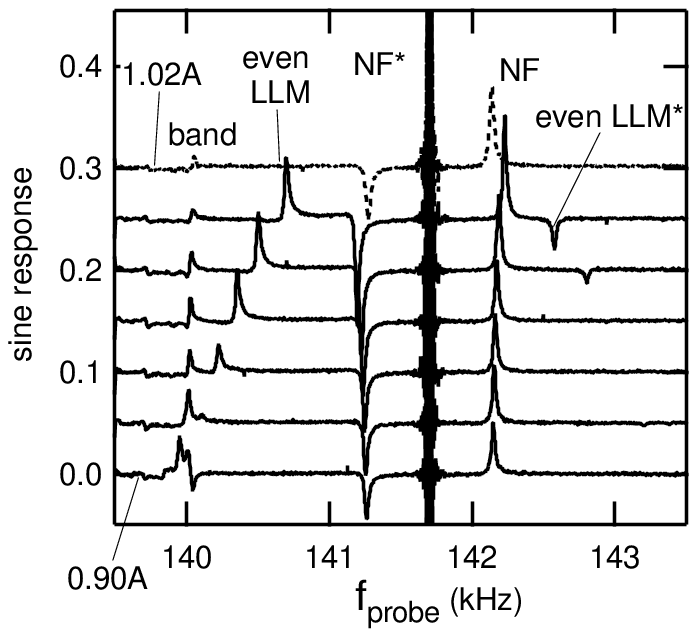}%
\caption{\label{fig:5}Sine response spectra for the IR laser repulsion transition for different values of IR laser current. From bottom to top, the IR laser current was increased from 0.90A($\sim $9mW)  to 1.02 A($\sim $18mW) with 0.02A steps. The pump frequency (ILM frequency) was fixed at 141.7kHz. The repulsion transition takes place between 1.00 and 1.02 A(dashed curve).  As the laser power increases, the even-LLM approaches the four-wave mixed peak NF$^*$. After the transition, the even-LLM disappears.}%
\end{figure}

Linear response curves versus IR laser currents are presented in Fig.~\ref{fig:5}. With increasing laser current, the even LLM frequency approaches the NF$^*$ peak. Just before the transition (1.00A), the NF is pulled toward the even-LLM. After the transition takes place (1.02A), the NF returns to the original position and the even-LLM returns to slightly above the band modes with very small peak height.

\subsection{Visible laser-induced ILM attraction}
\begin{figure}
\includegraphics{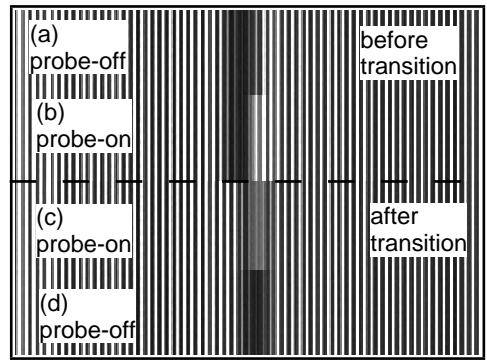}%
\caption{\label{fig:6}Images for visible laser attractive transition.  Only the impurity induced by the probe laser beam is used. Vertical white lines are cantilevers. The driver frequency is step decreased to reach the transition. (a) The ILM. (b) The probe laser is focused to the right side side of the ILM. (c) The driver frequency is decreased so at some point the ILM is attracted to the probe laser site. (d) Probe laser off.}%
\end{figure}

Figure~\ref{fig:6} demonstrates the attractive interaction produced by the visible laser impurity. In this case the IR laser is turned off and the power of the visible laser is increased to 1 mW. (Note the same laser was operated at 0.06mW in the previous IR laser experiment.) Instead of increasing the laser power further, we change the pump driver frequency (ILM frequency), because the effect of the visible laser is so weak the transition only takes place when the even-LLM frequency is placed at the symmetric position with respect to the NF peak. These modes change their frequencies as the driven ILM frequency is changed.

\begin{figure}
\includegraphics{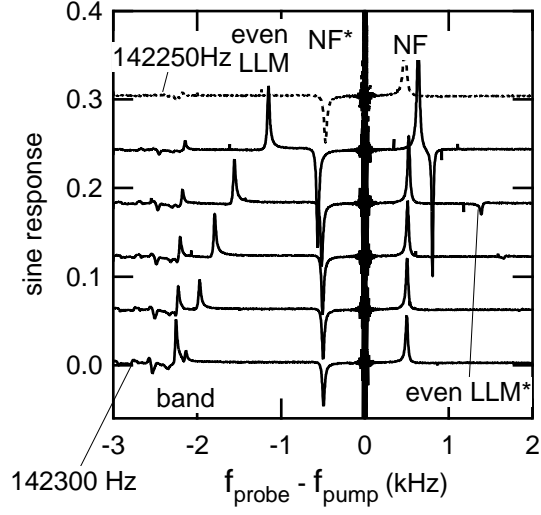}%
\caption{\label{fig:7}Sine response spectra for the visible-laser impurity attraction transition versus driver frequency. The driver frequency decreases from bottom to top:142.3kHz to 142.25 kHz with 10 Hz step. The visible probe laser is operated at the higher power of 1mW. When the even-LLM frequency approaches the four-wave mixed NF peak (NF$^*$), the transition occurs. After the transition, the even-LLM peak disappeared (top dashed curve).}%
\end{figure}

	We obtain very similar spectra to those found for the IR repulsion case. Figure~\ref{fig:7} presents the sine spectra for several driver frequencies. The even-LLM is made observable by the symmetry change induced by the visible laser near the ILM.  Its frequency approaches the ILM frequency as the driver frequency is decreased. In this figure the driver frequency decreases from bottom to top. The transition takes place when the even-LLM frequency is coincident with the NF$^*$.  After the transition, the even-mode disappears because the laser focus is now at the center of the ILM and lattice symmetry is recovered.

\section{\label{sec:simulation}Simulations and comparison with experiment}
\subsection{ILM repulsion and attraction transitions by impurities}
A lumped element model is used for the follow-on simulations with the equations of motion of the form:

\begin{eqnarray}
&&m_i \ddot x_i  + \frac{{m_i }}{\tau }\dot x_i  + k_{2O,i} x_i  + k_{4O} x_i^3\nonumber  \\ 
&&+ \sum\limits_j {k_{2I}^{(j)} \left( {2x_i  - x_{i + j}  - x_{i - j} } \right)}\nonumber   \\ 
&&+ k_{4I} \left\{ {\left( {x_i  - x_{i + 1} } \right)^3  + \left( {x_i  - x_{i - 1} } \right)^3 } \right\}\nonumber  \\ 
&&= m_i \alpha _{pump} \cos \Omega t + m_i \alpha _{probe} \cos \omega t  
\label{eq:1}
\end{eqnarray}

\noindent where $i$ is the site number of the cantilever, $m_i$ is the mass, $\tau $ is the relaxation time, $k_{2O,i}$ and $k_{4O}$ are harmonic and quartic onsite spring constant, $k_{2I}^{(j)}$ is the harmonic spring constant for the intersite connection up to $j$-th neighbor, and $k_{4I}$ is the quartic spring constant for the intersite connection. The right hand side is the driving term. Here to match experiment $\alpha _{pump}=1000$m/s$^2$ is the pump acceleration and $\Omega $ is the pump frequency.  The second term is for the probe driver at frequency $\omega $ and acceleration amplitude $\alpha _{probe}=0.01$m/s$^2$. The specific lattice parameters are listed in Table~\ref{table:1}. Those are determined by experimental observation and comparison with simulations. Numbers are the same as those in Ref.~\cite{9}.  Fixed boundary conditions are used and the total number of cantilevers is 100.

\begin{table*}
\begin{center}
\caption{\label{table:1}
Lumped oscillator parameters used in the simulations Values are the same as in Ref.~\cite{9}.
}
\begin{tabular}{ccccccc}
\hline
symbol & $m_i$ & $k_{2Oi}$ & $k_{2I}^{(j)}$ & $\tau $ & $k_{4O}$ & $k_{4I}$ \\
 & (kg)$\footnote{Upper row is for the longer cantilever, lower is for the shorter cantilever.}$ & (N/m)$\footnotemark[1]$ & (N/m)$\footnote{Listed from nearest neighbor to 6th nearest.}$ & (s) & (N/m$^3$) & (N/m$^3$) \\
\hline
value & $7.67\times 10^{-13}$ & 0.142277 & 0.0828453 & $8.75\times 10^{-3}$ & $1.0\times 10^8$ & $4.0\times 10^{10}$ \\
 & $6.98\times 10^{-13}$ & 0.168389 & 0.0308231 &  &  &  \\
 &  &  & 0.010831 &  &  &  \\
 &  &  & 0.00404721 &  &  &  \\
 &  &  & 0.00249521 &  &  &  \\
 &  &  & 0.000823741 &  &  &  \\
\hline
\end{tabular}
\end{center}
\end{table*}

\begin{figure}
\includegraphics{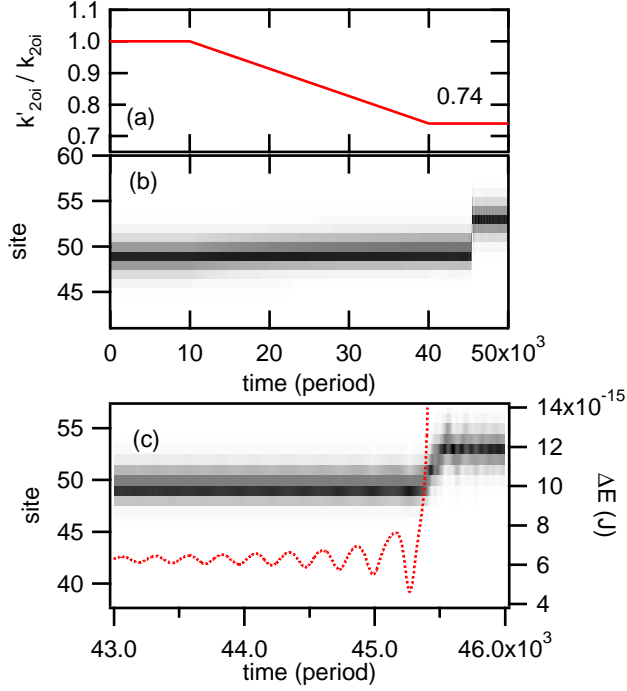}%
\caption{\label{fig:8}Time dependent simulations of the repulsion transition by the soft impurity. The ILM hops to a new position via the auto-oscillation of its lateral motion. The driver amplitude $\alpha _{pump}=1000$m/s$^2$  and the pump driver frequency   $(F-f_T)/f_{BW}=0.454$ ($F$=139kHz) are fixed, where $f_T=137.1$kHz is the band top frequency and $f_{BW}=4.1$kHz.  (a) The strength of the harmonic impurity in this anharmonic lattice is gradually decreased till the threshold $k'_{2O,i}/k_{2O,i}=0.74$. (b) The impurity is located at site 47, and the ILM is initially at 49. The ILM is repelled to site 53. After the impurity strength reaches the threshold, but transition takes place at a later time.  (c) Magnified picture of  panel (b) at around the transition. Before the transition, the lateral motion of the ILM increases with time. Dashed curve is the difference energy between the two sites  $\Delta E=E_{50}-E_{48}$. }%
\end{figure}

Figure~\ref{fig:8} summarizes the repulsion transition as determined from simulations.  A harmonic impurity located to one side of the ILM repels it when the impurity ratio $k'_{2O,i}/k_{2O,i}$ exceeds a threshold, where  $k_{2O,i}$ and $k'_{2O,i}$  are the onsite linear spring constant for the regular and impurity sites, respectively.  When the cantilever nonlinearity is positive,  $k'_{2O,i}/k_{2O,i}<1$ is required to repel the ILM. For this particular case, the threshold is 0.74. The harmonic spring constant is changed as a function of time as shown in Fig.~\ref{fig:8}(a). Figure~\ref{fig:8}(b) presents the energy density plot versus time and shows the displaced ILM after the spring constant ratio reaches the threshold. The energy difference between sites 48 and 50 is displayed in Fig~\ref{fig:8}(c) with the magnified density plot around the transition. Clearly the transition is due to the amplification of lateral oscillation with time, not by a smooth rectified translational motion of the ILM.

\begin{figure}
\includegraphics{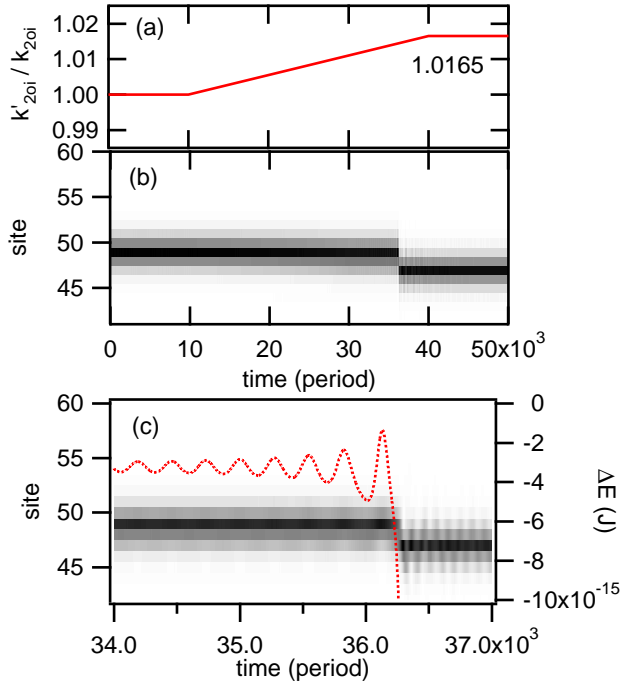}%
\caption{\label{fig:9}Time dependent simulations of the attractive ILM transition produced by the hard impurity. The ILM hops to a new position via auto-oscillation of its lateral motion. The driving conditions are the same as the soft impurity case. (a) The strength of the harmonic impurity is gradually increased till the threshold  $k'_{2O,i}/k_{2O,i}=1.0165$. (b) The impurity is located at site 47, and the ILM is initially at 49. The ILM is attracted to site 47. After the impurity strength reaches the threshold, but transition takes place at a later time.  (c) Magnified picture of  panel (b) at around the transition. Before the transition, the lateral motion of the ILM increases with time. Dashed curve is the difference energy between the two sites  $\Delta E=E_{50}-E_{48}$.}%
\end{figure}

For the attractive transition by the hard impurity, the threshold of the hard impurity is $k'_{2O,i}/k_{2O,i}=1.0165$ for the same driving condition as the soft impurity case. The impurity spring constant is increased with time as shown in Fig.~\ref{fig:9}(a). The hard impurity is located at site 47, the same site as for the soft impurity case, but now the ILM is attracted to it as shown in Fig.~\ref{fig:9}(b). The energy difference between sites 48 and 50 again oscillates before the transition as shown in Fig.~\ref{fig:9}(c). This signature indicates that the transition mechanism is the same as for the soft impurity. Note that the energy difference is mostly negative in this case while it is mostly positive in Fig.~\ref{fig:8}(c). This observation implies that the center of the ILM vibration is slightly shifted towards the impurity in Fig.~\ref{fig:9}(c), while it is shifted away from the impurity in Fig.~\ref{fig:8}(c).

\subsection{Linear response spectra}
\begin{figure}
\includegraphics{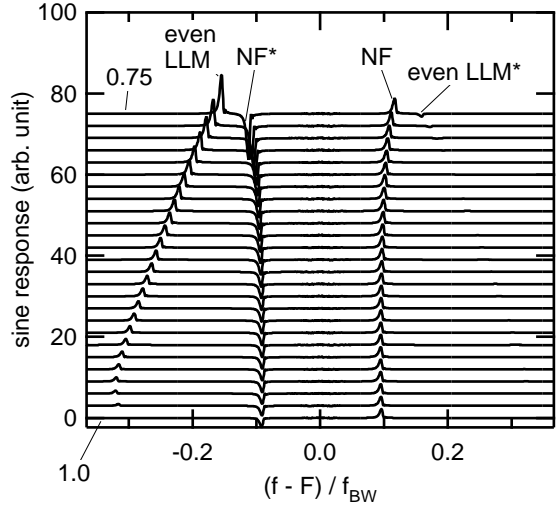}%
\caption{\label{fig:10}Simulated sine response spectra for the repulsive ILM transition versus impurity strength. Impurity strength increases from bottom to top. When the even LLM and the NF$^*$ merge, the transition takes place. The impurity strength increases from $k'_{2O,i}/k_{2O,i}=1.00$(pure) to 0.75 (bottom to top) The transition threshold is 0.74. By increasing the impurity strength, the even LLM is activated and its peak height increases.}%
\end{figure}

To find the movement of the linear modes in the spectrum, the probe response spectra are calculated the same way as with experiments. The uniform probe driver is used. Displacement at next short cantilever site of the ILM is analyzed by multiplying the response by $\sin \omega t$ or $\cos \omega t$, like a lock-in amplifier, and then time-averaging for sine and cosine responses. Two sets of simulations with opposite probe phase are made, and then subtracted to eliminate the large amplitude ILM vibration from consideration.  The resultant sine and cosine spectra are very similar to the experiments shown in Fig.~\ref{fig:4}. Only the sine components are shown in Fig.~\ref{fig:10} for various impurities, from 1.00(pure) to 0.75 (before the bifurcation). As the bifurcation point is approached, NF$^*$ and the even LLM move close to each other, and both peaks are enhanced. The simulation and experimental results are in good agreement, demonstrating that the hopping transition is due to the softening of the cantilever resonance frequency by the IR laser. (A decrease in the Young's modules of silicon nitride with heating is reported by Bruls et al.~\cite{24}.) 

The repulsion is triggered by the interaction between the NF and the even LLM. The enhancement of the response peaks in Fig.~\ref{fig:10} can be explained as follows. The signal at the upper sideband peak of the NF generates four-wave mixing signal at the NF$^*$ frequency. If the NF$^*$ coalesces with the even-LLM mode frequency, the four-wave mixing signal of the NF resonates with the even-LLM. This signal is four-wave mixed again and returns to the NF frequency. The total loop gain can be far larger than without the coalescence. If the loop gain overcomes the loss, the coupled NF and even-LLM grow without any driving source. Thus the translational vibration of the ILM diverges. This scenario is consistent with Figs.~\ref{fig:8}(c) and \ref{fig:9}(c), where the lateral motion is growing with time.

Clearly the even-LLM is a key element. The difference frequency between the driven ILM and the LLM is the ILM lateral vibration frequency. There could be two ways to approach this lateral transition of the ILM:(1) continuously soften the lateral vibration frequency so that the final translation is described by a frozen vibration of the lateral motion, or (2) a diverging amplitude of the lateral motion at a fixed frequency that overcomes the pinning barrier. The second case is observed in our experiments. When decreasing the driver frequency, the even-LLM frequency continuously decreases but the transition takes place when the even LLM coalesces with NF*, before the even-LLM completely softens. In contrast for the first possibility the even-LLM peak would pass through the NF* and continuously approach the central ILM peak at the transition.

The hopping mechanism for the visible attraction of the ILM is the same as for the IR laser repulsion case, e.g., generation of the translational vibration at the symmetric frequency condition of the two linear modes. An experimental detail is that because of the very weak hardening of the spring constant, the translation can be seen only by scanning the driver frequency.  At the symmetric even-LLM and NF condition, the transition takes place even if the hardening effect is very small. In our case the effect is so small that the translational motion happens only by controlling the ILM frequency to tune the even-LLM and NF to satisfy the nonlinear mixing condition $f_{NF}+f_{even-LLM}=2f_{ILM}$.  In simulations, we have succeeded in reproducing the translational oscillation plus the parametric effect with the impurity level   $k'_{2O,i}=1.0165k_{2O,i}$, while the experimental value is much smaller: $k'_{2O,i}=1.0004k_{2O,i}$. This difference is most likely due to our approximate  simulation model. 

It should be emphasized these hopping transitions appear similar to the lower bifurcation transition of the AR region without the impurity when the pump frequency is used as a control parameter.\cite{21} In that case NF$^*$ coalesces and interacts with the top band mode at the transition.\cite{21} In our present picture this would correspond to the NF and the band mode frequencies being symmetrically placed with respect to the ILM frequency.  So both for intrinsic and impurity cases the transition is controlled by the NF and the other linear mode of interest being symmetrically placed about the ILM frequency. They both satisfy  $f_{NF}+f'=2f_{ILM}$ where $f'$ is the frequency of the even-LLM or the band mode. It is a parametric process that drives the auto oscillation mechanism in both cases.

\subsection{Harmonic lattice with ILM-like mode plus an impurity}
To examine the difference between repulsion and attraction of the ILM, we tested a linear harmonic lattice eigenvector calculation by constructing an ILM-like impurity mode using inter-site spring constant changes. The impurity spring constants were estimated from the ILM eigenvector for a pure case. In this way the increment of the spring constant from the regular intersite spring  $k_{2I}^{(1)}$  is proportional to the square of the difference displacement $(x_i-x_{i-1})^2$, where $x_i$  is the eigenvector. The particular set of values used for simulations are $\Delta k_{2I,i}=k_{ILM}($ ..., 0.003, 0.017, 0.104, 0.368, 1.00, 1.00, 0.368, 0.104, 0.017, 0.003, ....) where the ILM center is between two 1.00s, and a parameter adjusting the ILM effect $k_{ILM}=0.2k_{2I}^{(1)}$ is used.  The intersite springs are changed from those in Eq.~\ref{eq:1} by using this parameter set so $k_{2I}^{(1)}\rightarrow k_{2I}^{(1)}+\Delta k_{2I,i}$. The onsite impurity, which is either soft or hard, is placed at the next short cantilever site just as before, by changing the onsite spring constant at that particular site. The final linearized equation for calculating the eigenvector is 

\begin{eqnarray}
m_i \ddot x_i  + k_{2Oi} x_i  + \sum\limits_j {k_{2I,i}^{(j)} \left( {2x_i  - x_{i + j}  - x_{i - j} } \right)}  = 0
\label{eq:2}
\end{eqnarray}

\noindent The eigenvectors are then calculation in the usual way, with the ILM-like mode the highest frequency one.

\begin{figure}
\includegraphics{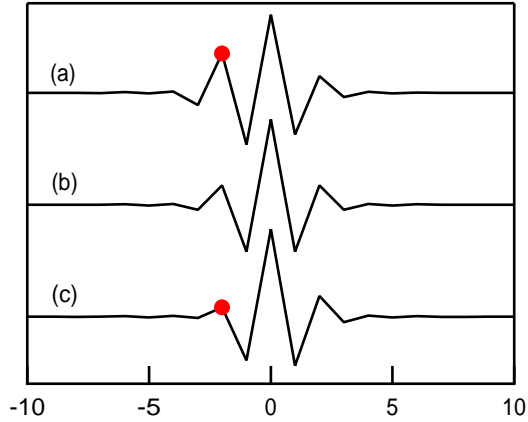}%
\caption{\label{fig:11}Harmonic eigenvector for an ILM-like mode with a nearby onsite impurity. Impurity location is indicated by the solid dot. (a) Hard impurity $k'_{2O,i}/k_{2O,i}=1.2$. (b) Pure ILM-like mode. (c) Soft impurity $k'_{2O,i}/k_{2O,i}=0.6$.}%
\end{figure}

Figure~\ref{fig:11} shows the result of calculation for (a) the hard impurity, (b) no on site impurity, and (c) soft impurity.  A symmetric eigenvector is obtained for the pure case as expected. For the hard impurity case shown in (a), the ILM-impurity mode has a larger amplitude at the impurity site and a smaller amplitude on the opposite side, indicating that the ILM center gravity is shifted towards the impurity. On the other hand, for the soft impurity case in (c), the mode has smaller amplitude at the impurity site and larger amplitude at the opposite side, indicating that the ILM center gravity has moved away from the impurity.  These results are consistent with the experiment and also with the nonlinear simulations.

Here, we have used relatively stronger impurities: $k'_{2O,i}/k_{2O,i}=1.2$ for the hard case  and $k'_{2O,i}/k_{2O,i}=0.6$ for the soft case to see the effects more clearly. In the real system, the nonlinear effect on intersite springs should change as the eigenvector changes, so it is expected that the effect of the impurities on the eigenvector shown in Fig.~\ref{fig:11} will be smaller than for the nonlinear simulations, because the intersite impurity spring constants are fixed at the pure ILM value. Thus, the direction of motion is the same in this linear calculation and the nonlinearity will only increase the movement. 

\subsection{Mechanism of cantilever hardening}
Here we review the possible mechanisms that can increase the cantilever vibration frequency under visible laser radiation. One is due to a light force.\cite{5} But we find that for our situation the estimated spring constant increase by this mechanism is very small   $\sim 1 \times 10^{ - 9}\%$ at  $P=$1.4mW, too small to be measured with our setup.

Photostriction has been reported in silicon \cite{25} while expansion was observed in germanium \cite{26}. Cantilevers made from silicon have been studied experimentally\cite{1} and theoretically\cite{2} for gas or chemical sensors, which bend cantilevers under light irradiation. Shrinkage or expansion can be determined by the sign of the pressure dependence of the band gap. A positive sign of the pressure dependence $\partial E/\partial p$ means positive strain under photo irradiation. Unfortunately, there is no report on the pressure dependence of the band gap of silicon nitride.  

On the other hand, it has been shown theoretically and experimentally that a total surface stress of upper and lower surfaces of a cantilever can modify the resonance frequency, while the difference surface stress between these sides gives rise the bending. \cite{27,28,29,30} To explain our positive frequency shift by this effect, a negative total surface stress would be expected. If the pressure dependence of the band gap is negative, the blue shift by visible laser irradiation can be explained. 

Another possibility, and the one we favor, is stress generation by photo-excited excess carriers. The visible laser (658 nm, 1.9eV) has larger photon energy than that of the IR laser (808 nm, 1.54eV). The band gap of the silicon nitride is from 1.8 to 5.1eV, depending on stoichiometry. Since the visible laser causes hardening while IR laser causes softening, this difference is most likely due to photo-excitation of electrons and holes in the nitride cantilever. 

No doubt if we used a much higher power visible laser there would be sufficient heat generated that the temperature effect would overcome the photo-carrier effect. In this case, the visible laser would repel the ILM. Since the photon energy of the IR laser is smaller than the band gap of the SiN, the radiation may not be absorbed strongly by the cantilever; however, the power of the IR laser was much larger than for the visible case. In addition, the silicon substrate also absorbs the IR radiation because its band gap 1.1 eV is smaller than the photon energy. Since cantilevers and overhangs rest on the silicon substrate, and the IR laser is softly focused, the heat absorbed by the silicon may warm up the local cantilevers. 

\section{Conclusions}
We have found that visible laser radiation hardens the silicon nitride cantilever, most likely due to photo-excited excess carriers, so that its impurity mode attracts the ILM. The attractive hopping transition of the AR-ILM by the visible laser and the repulsive transition induced by the IR-laser observed in the 1-D micromechanical cantilever array have been examined experimentally and numerically. At the transition points for both kinds of impurities, the NF and the even-LLM are symmetrically positioned with respect to the ILM frequency so that the transition occurs when $f_{NF}+f_{even-LLM}=2f_{ILM}$. This signature condition produces parametric excitation of the NF and even-LLM, drives the lateral motion of the ILM, and this amplified lateral motion causes the translational hopping transition. Interestingly numerical simulations with a harmonic model containing impurities correctly give the sign of the observed interaction but our nonlinear simulations demonstrate that the magnitude of the effect is enhanced in a nonlinear lattice. Finally, casting these results in a larger framework the laser-induced impurity and its impurity mode simply provide a way to connect the ILM to the external world so that its intrinsic translational dynamics can be explored. Should a different technique be applied to tickle a driven ILM in a general 1-D nonlinear lattice with onsite and intersite coupling, to explore its translational transition, then the intrinsic NF and even-LLM excitations described here would still play the same defining dynamical role.

\begin{acknowledgments}
M.S. was supported by JSPS-Grant-in-Aid for Scientific Research No. 25400394. A.J.S. was supported by Grant NSF-DMR-0906491. M.S. thanks Dr. M. Kimura at University of Shiga Prefecture for fruitful discussions. A.J.S. acknowledges the hospitality of San Diego State University and Kanazawa University, where some of this work was completed. 
\end{acknowledgments}


\providecommand{\noopsort}[1]{}\providecommand{\singleletter}[1]{#1}%

\end{document}